
\magnification=1200
%
%
\def\half{{1\over 2}}

\def\vp{{\varphi}}
\def\p{{\partial}}
\tolerance=5000
\footline={\ifnum\pageno>1
       \hfil {\rm \folio} \hfil
    \else \hfil \fi}

\overfullrule=0pt 
\baselineskip=18pt
\raggedbottom
\centerline{\bf A $W$-String Realization of the Bosonic String}
\vskip 12pt
\centerline{Nathan Berkovits, Michael Freeman and Peter West}
\vskip 12pt
\centerline{Maths Dept., King's College, Strand, London, WC2R 2LS, United
Kingdom}
\centerline{e-mail: udah101@oak.cc.kcl.ac.uk}
\vskip 12pt
\centerline {KCL-TH-93-15}
\vskip 12pt
\centerline {November 1993}
\vskip 24pt
\centerline {\bf Abstract}
It has recently been shown that the ordinary bosonic string
can be represented by a special background of
N=1 or N=2 strings. In this paper, it will be shown that the bosonic string
can also be represented by a special background of $W$-strings.

\noindent
\vfil\eject
\vskip 12pt

By adding fermionic gauge fields
to the bosonic string, it was recently shown
that an N=1 superconformal algebra could
be constructed [1],
and that the cohomology of the
corresponding
N=1 BRST charge
coincides with the cohomology of the original bosonic string [2,3].
Furthermore, it was shown that the N=1 prescription for calculating
scattering amplitudes with this special choice of matter fields
produces the usual bosonic string amplitudes [1,3], allowing one to view the
bosonic string as a background of N=1 strings.
It was speculated first by Vafa [1], and later by others [2,4],
that the bosonic string might similarly be
represented by a special background of $W$-strings.

In this paper, it will be shown
that by adding bosonic gauge
fields to the bosonic string, a $W_3$ algebra can be constructed, and
the cohomology of the corresponding $W_3$ BRST charge is precisely the
cohomology of the ordinary bosonic string.
If it is assumed that $W$-moduli can be safely
ignored in this special background,
the resulting scattering amplitudes are just
those of the ordinary bosonic string.
These results easily generalize to the case of
$W_{2,s}$ strings which contain one spin 2 generator and one spin $s$
generator.

\vskip 15pt
Over the past few years, it was shown that for a particular
representation of the $W_3$ algebra, the corresponding
$W_3$ string theory
has a cohomology containing Ising model primary fields and has tree-level
scattering amplitudes which contain Ising model correlators in their
integrands [5].
These results suggest that this $W_3$ string can be thought of as a
non-critical bosonic string coupled to a $c=\half$ Ising model.
The matter fields in this particular representation of the $W_3$
string consist of a scalar boson $\varphi$ with background charge
$\sqrt{49\over 8}$ and $d$ scalar bosons $x^\mu$ ($\mu =0$ to $d-1$) whose
background charge $\alpha^\mu$ satisfies $\alpha^\mu\alpha_\mu=(d-25\half)
/12$.

It will be shown in this paper that for the choice $d=27$ and
$\alpha^\mu=(0,...,0,1/\sqrt{8})$, the cohomology of the BRST charge
for this $W_3$ string, when expressed in terms of a
new set of fields, corresponds to the cohomology
of an ordinary bosonic string.
This will be proven by showing that the effects of
$x^{26}$ and $\varphi$ are precisely
canceled by the spin $(3,-2)$ $W_3$ ghosts,
$d$ and $e$. In other words, the $x^{26}$ and $\varphi$ fields can be
completely gauged
away using the $W_3$-transformations, leaving only the 26 $x^\mu$'s
of the ordinary bosonic string. By replacing the first
26 $x^\mu$'s by any other $c=26$ system, this same proof can be used to
show that any $c=26$ matter system can be represented by a $W_3$ string.

The proof will consist of finding a field redefinition such that
the BRST charge for this representation of the $W_3$ string takes the
simple form:
$$Q= Q_B+ \int dz f G\eqno(1)$$
where $Q_B=\int dz (cT_B +c \partial
c b)$ is the BRST charge for the ordinary bosonic
string (or any other matter system whose central charge is 26) and $fG$ is
a non-minimal
term constructed out of $x^{26}, \varphi, d,$ and $e$ in such a way
that ($f$,$g$) are a conjugate pair of free fermions and ($F,G$) are
a conjugate pair of free bosons. Since ($f,g,F,G$) appear only in this
non-minimal term, the ``quartet'' argument of Kugo and Ojima can be used
to prove that the cohomology of $Q$ is equivalent to the cohomology of
$Q_B$.

The BRST charge for this particular representation of the
$W_3$ string is given by

$$
Q=\int dz [ c(T^m + {1\over2} T^{gh}) + e(W^m + {1\over 2} W^{gh})],\eqno (2)
$$
where
$$
T_m  = T_\varphi + T_x ,
\quad
T_\varphi  = -{1\over 2}{(\partial\varphi)}^2 -
  \sqrt{49\over 8} \partial^2\varphi,\quad
T_x  = T_B-\half(\p x^{26})(\p x^{26})-{i\over \sqrt{8}}\p^2 x^{26},
$$
$$
T^{gh}   = - 2 b\,\partial c - \partial b\,c - 3d\, \partial e -
2 \partial d\, e , \eqno(3)
$$
$$
W_m  =- {2i\over \sqrt{261}}\left[{1 \over 3} { (\partial \varphi)}^3 +
  \sqrt{{49}\over 8} \partial\varphi \partial^2 \varphi
  +{1 \over 3} {{49}\over 8} \partial^3 \varphi +
  2 \partial\varphi T_x + \sqrt{{49}\over 8} \partial T_x\right], \eqno(4)
$$
$$
W^{gh}  = - \partial d\, c - 3 d\, \partial c -{8\over 261} [\partial
(b\,e\, T^m) + b\,\partial e\, T^m]
$$
$$
 + {25\over 1566} (2 e \partial^3 b + 9 \partial e\, \partial^2 b
  + 15 \partial^2 e\, \partial b + 10 \partial ^3 e\, b). \eqno(5)
$$
The first step in simplifying $Q$ is to use the field redefinition
of reference [6]:
$$\tilde c=c+{7\sqrt{58}i\over174}\p e-{8\over 261}b\p e e-
{4\sqrt{29}i\over 87} \p\vp e,\eqno(6)$$
$$\tilde d=d+{7\sqrt{58}i\over174}\p b-{8\over 261}\p b b e+
{4\sqrt{29}i\over 87} \p\vp e,\quad
\tilde\vp=\vp-
{4\sqrt{29}i\over 87} b, $$
which allows $Q$ to be written in the simpler form:
$$Q = Q_0 + Q_1 \eqno(7)$$
where
$$Q_0=\int dz [\tilde c (T_B-\half(\p x^{26})^2
-{i\over \sqrt{8}}\p^2 x^{26} +
T_{\tilde\varphi}+T_{\tilde d e})+\tilde c
\partial \tilde c b],$$
$$Q_1=\int dz \left[-{4\sqrt{29}i \over 261}  e \left(
2(\p\tilde\varphi)^3+{42\over\sqrt{8}}\p^2\tilde\vp\p\tilde\vp+
{19\over 4}\p^3\tilde\vp+9\p\tilde\vp\tilde d\p e+
{21\over\sqrt{8}}\p\tilde d\p e\right)\right]\eqno(8).
$$
In $Q_0$, the stress tensors for the redefined fields are the same as those
of the original fields.

The next step in simplifying $Q$ is to define new fields [7]
$$\phi_1= -3\rho -i\sqrt{8}\tilde\varphi,\quad
\phi_2= -i\sqrt{8}\rho +3\tilde\vp\eqno(9)$$
where $\rho$ comes from bosonizing the $W_3$ ghosts in the standard way:
$$\tilde d=e^{-i\rho},\quad  e=e^{i\rho}. \eqno(10)$$
Since $\phi_1$ and $\phi_2$ have free-field OPE's, it is easy to show that
$$Q_0=\int dz [\tilde c (T_B-\half(\p x^{26})^2
-{i\over \sqrt{8}}\p^2 x^{26} +
T_{\phi_1}+T_{\phi_2})+\tilde c
\partial \tilde c  b], \eqno(11)$$
where
$$T_{\phi_1}= -\half(\p\phi_1)^2 -{i\over 2}\p^2\phi_1,\quad
T_{\phi_2}=-\half(\p\phi_2)^2-{1\over\sqrt{8}}\p^2\phi_2.\eqno(12)$$

$Q_1$ can be checked to take the following form when expressed in terms
of $\phi_1$ and $\phi_2$ [7]:
$$
Q_1 = \int dz
e^{\sqrt{8}\phi_2} e^{-4i\phi_1}\partial^3 e^{i\phi_1} \eqno(13)
$$
where we have dropped an overall irrelevant factor of $-i\sqrt{58}/1044$
from the expression for $Q_1$.
This can be simplified further by defining a new field $\hat\phi_1$ in
terms of $\phi_1$ by means of the relations [7]
$$
\eqalign{
e^{i\hat\phi_1} &= i \partial\phi_1 e^{-i\phi_1} \cr
i\partial\hat\phi_1 e^{-i\hat\phi_1} &= - e^{i\phi_1}.
}\eqno(14)
$$
By taking repeated operator products of $e^{i\hat\phi_1}$ with itself it can
be shown that,
in terms of $\hat\phi_1$ and $\phi_2$, $Q_1$ takes the simple form [7]
$$Q_1=\int dz e^{3i\hat\phi_1} e^{\sqrt{8}\phi_2}. \eqno(15)$$
Furthermore, because all fields satisfy free-field OPE's, $Q_0$
remains of the form
$$Q_0=\int dz [\tilde c (T_B-\half(\p x^{26})^2
-{i\over \sqrt{8}}\p^2 x^{26} +T_{\hat\phi_1}+T_{\phi_2})+\tilde c
\partial \tilde c  b], \eqno(16)$$
where $T_{\hat\phi_1}$ is the same stress tensor as $T_{\phi_1}$.

The last step in simplifying $Q$ is to combine $x^{26}$, $\hat\phi_1$,
and $\phi_2$ into a conjugate pair of free fermions $(f,g)$
and a conjugate pair of free bosons $(F,G)$.
This is done by defining
$$f=e^{iw},\quad g=e^{-i w},\quad
F=\p (e^{-iy}) e^{-\sigma}\quad
G=e^{iy} e^{\sigma}, \eqno(17)$$
where
$$w= \hat\phi_1-{i\over\sqrt{2}}\phi_2-{1\over\sqrt{2}}x^{26},\quad
y= \hat\phi_1-{i\over\sqrt{2}}\phi_2+{1\over\sqrt{2}}x^{26},\quad
\sigma=i\hat\phi_1+\sqrt{2}\phi_2. \eqno(18)$$
The coefficients in equation (18) have been chosen such that
$$Q_1=\int dz f G. \eqno(19)$$
Furthermore, since
$f$, $g$, $F$, and $G$ are bosonized in the standard way for free
fermions and free bosons, it is
easy to show that
$$Q_0=\int dz \left[\tilde c
\left(T_B + \half(\p f g-f\p g) +\half (\p F G- F\p G)\right)
+ \tilde c\partial \tilde c  b\right].\eqno(20)$$
Finally, by defining
$$\hat f=f+\tilde c\p F+\half\p \tilde c F
,\quad\hat G=G+\tilde c \p g+\half\p\tilde c g,\quad
\hat b=b+\half F\p g-\half g\p F,\eqno(21)$$
the BRST charge for this particular representation of the $W_3$ string
takes the form
$$Q=Q_0 +Q_1=\int dz[(\tilde c T_B-\tilde c\p\tilde c\hat b)  +\hat f\hat G].
\eqno(22)$$

Since $(\hat f,g,F,\hat G)$ only appear in the last term of equation (22),
the ``quartet'' argument of Kugo and Ojima can be used to
prove that all dependence of physical states
on these extra fields can be gauged away,
and therefore the cohomology
of $Q$ is simply the cohomology of the BRST charge for the bosonic string,
$Q_B=\int dz (\tilde c T_B +\tilde c \partial\tilde c \hat b)$.
Furthermore, since the field redefinitions of equations (6), (9), (14)
and (21) preserve all OPE's,
correlation functions of physical vertex operators of this $W_3$ string
are the same as correlation functions of the corresponding physical
vertex operators of the bosonic string. Therefore, assuming that the role of
$W$-moduli can be ignored, the resulting scattering
amplitudes are just those of the ordinary bosonic string.

Note that the ``quartet'' argument of Kugo and Ojima depends
on the assumption that all
$W_3$-string
states can be
constructed from the vacuum using
creation modes of the $(\hat f,g,F,\hat G)$ fields and
creation modes of the
bosonic string matter and ghost fields.
Although this includes all states which are
normalizable using
the standard free-field
norm of the $(\hat f,g,F,\hat G)$ fields, it does not include
all states which are normalizable using the original norm
defined for the $(x^{26},\phi,d,e)$ fields.
For example, there are many states
which are normalizable in terms of the original norm which can not be
constructed from a vacuum using only creation modes
of the $(\hat f,g,F,\hat G)$ fields and
creation modes of the
bosonic string matter and ghost fields. In fact of all the physical vertex
operators described in reference [6], only the operator $\tilde
cV$ for $k^{26}=0$ can be constructed in this way.
For this reason, the cohomology of $Q$
which was found in references [6,8]
using normalizable states built out of the
$(x^{26},\phi,d,e)$ fields
is much bigger than the cohomology found in this paper
using normalizable states built out of the $(\hat f,g,F,\hat G)$ fields.

This $W_3$ string realization of the bosonic string is easily generalized
to the $W_{2,s}$ string. In reference [7] it was explained how, starting from
the bosonic field $\tilde\varphi$ and bosonized ghost $\rho$ of the
$W_{2,s}$ string, we can define new fields
$$
\eqalign{
\phi_1 &= -s \rho -i \sqrt{s^2-1} \tilde\varphi \cr
\phi_2 &= -i \sqrt{s^2-1} \rho + s \tilde\varphi
}\eqno(23)
$$
in terms of which the non-trivial part $Q_1$ of the BRST charge
becomes
$$
Q_1 = \int  dz\, e^{\sqrt{s^2-1} \phi_2} e^{-i(s+1)\phi_1}
\partial^s e^{i\phi_1}. \eqno(24)
$$
We can rewrite this expression in terms of a field $\hat\phi_1$
defined by equation (14), to obtain
$Q_1 =\int dz\exp\{\sqrt{s^2-1}\phi_2 - i s \hat\phi_1\}$, and then the
quartet mechanism can be used to show that the cohomology of the BRST
charge is the same as that of the bosonic string.

\vskip 15pt

In this paper, we have found a representation of the $W_3$ string
that reduces to the bosonic string; namely, its physical states and
scattering amplitudes are those of the bosonic string. Thus the bosonic
string can be viewed as a special
background of the $W_3$ string, as well as the
N=1 and N=2 strings. It would seem likely in view of these results that
the bosonic string can be embedded into any string which is based on a
symmetry which includes the Virasoro algebra.

Although this might appear surprising, it is possible that
the strong consistency conditions of string theory inevitably
force a string to be the bosonic string once it has c=26 matter
tensored with an appropriate choice of gauge fields. A potential
analogy worth bearing in mind is the
situation that occurs in the theory of
non-linear realizations; given a classical theory
invariant under a rigid
symmetry group, we can promote it to be invariant under any larger
symmetry group which contains the original group by introducing
the appropriate Goldstone bosons [9].

\vskip 30pt
\centerline {\bf Acknowledgement}
\par
We wish to thank Jos\'e Figueroa-O'Farrill, Chris Hull, Martin
Ro\v cek, Kelly Stelle and
Cumrun Vafa for discussions.

\vskip 20pt
\centerline{\bf References}

\item{[1]} N. Berkovits and C. Vafa, ``On the uniqueness of string
theory,'' preprint HUTP-93/A031, KCL-TH-93-13, hep-th/9310129.
\item{[2]} J. M. Figueroa-O'Farrill, ``On the universal string theory,''
preprint QMW-PH-93-29, hep-th/9310200.
\item{[3]} H. Ishikawa and M. Kato, ``Note on N=0 string as N=1 string,''
preprint UT-Komaba/93-23, hep-th/9311139.
\item{[4]} Chris Hull, private communication.
\item{} Kelly Stelle, private communication.
\item{} H. Lu, C. N. Pope, X-J. Wang and S. C. Zhao, ``Critical
and non-critical $W_{2,4}$ strings,'' preprint CTP TAMU-70/93,
hep-th/9311084.
\item{[5]} For a review, see P. West, ``A review of $W$ strings,'' preprint
G\"oteborg-ITP-93-40, hep-th/9309095.
\item{[6]} H. Lu, C. N. Pope, S. Schrans and X-J. Wang,
Nucl. Phys. B408 (1993) 3.
\item{[7]} M. Freeman and P. West, ``Parafermions, $W$ strings and their
BRST charges,'' preprint KCL-TH-93-14.
\item{[8]} P. West, Int. J. Mod. Phys. A8 (1993) 2875.
\item{} H. Lu, B. E. W. Nilsson, C. N. Pope, K. S. Stelle and P. West,
Int. J. Mod. Phys. A8 (1993) 4071.
\item{[9]} S. Coleman, J. Wess and B. Zumino, Phys. Rev. 177 (1969) 2239.
\end